\documentclass[aps,prl,twocolumn,showpacs,superscriptaddress]{revtex4}
\usepackage{amssymb}
\usepackage{amsfonts}
\usepackage{graphicx}
\usepackage{CJK}
\usepackage{indentfirst}
\usepackage{amsmath}
\usepackage{epstopdf}
\begin{document}


\title{Multiple gaps revealed by low temperature specific heat in the 1111-type CaFe$_{0.88}$Co$_{0.12}$AsF single crystals}

\author{Jianan Chu}
\affiliation{State Key Laboratory of Functional Materials for
Informatics, Shanghai Institute of Microsystem and Information
Technology, Chinese Academy of Sciences, Shanghai 200050,
China}\affiliation{CAS Center for Excellence in Superconducting
Electronics(CENSE), Shanghai 200050, China}\affiliation{University
of Chinese Academy of Sciences, Beijing 100049, China}

\author{Teng Wang} \affiliation{State Key
Laboratory of Functional Materials for Informatics, Shanghai
Institute of Microsystem and Information Technology, Chinese Academy
of Sciences, Shanghai 200050, China}\affiliation{CAS Center for Excellence in Superconducting
Electronics(CENSE), Shanghai 200050, China}\affiliation{School of Physical Science and Technology, ShanghaiTech University, Shanghai 201210, China}

\author{Yonghui Ma}
\affiliation{State Key Laboratory of Functional Materials for
Informatics, Shanghai Institute of Microsystem and Information
Technology, Chinese Academy of Sciences, Shanghai 200050,
China}\affiliation{CAS Center for Excellence in Superconducting
Electronics(CENSE), Shanghai 200050, China}\affiliation{University
of Chinese Academy of Sciences, Beijing 100049, China}

\author{Jiaxin Feng}
\affiliation{State Key Laboratory of Functional Materials for
Informatics, Shanghai Institute of Microsystem and Information
Technology, Chinese Academy of Sciences, Shanghai 200050, China}\affiliation{CAS Center for Excellence in Superconducting
Electronics(CENSE), Shanghai 200050, China}
\affiliation{University of Chinese Academy of Sciences, Beijing 100049, China}

\author{Lingling Wang}
\affiliation{State Key Laboratory of Functional Materials for
Informatics, Shanghai Institute of Microsystem and Information
Technology, Chinese Academy of Sciences, Shanghai 200050,
China}

\author{Xuguang Xu} \affiliation{School of Physical Science and Technology, ShanghaiTech University, Shanghai 201210, China}

\author{Wei Li}
\affiliation{State Key Laboratory of Surface Physics and Department of Physics, Fudan University, Shanghai 200433, China}
\affiliation{Collaborative Innovation Center of Advanced Microstructures, Nanjing 210093, China}

\author{Gang Mu}
\email[]{mugang@mail.sim.ac.cn} \affiliation{State Key Laboratory of
Functional Materials for Informatics, Shanghai Institute of
Microsystem and Information Technology, Chinese Academy of Sciences,
Shanghai 200050, China}\affiliation{CAS Center for Excellence in Superconducting
Electronics(CENSE), Shanghai 200050, China}

\author{Xiaoming Xie}
\affiliation{State Key Laboratory of Functional Materials for
Informatics, Shanghai Institute of Microsystem and Information
Technology, Chinese Academy of Sciences, Shanghai 200050, China}\affiliation{CAS Center for Excellence in Superconducting
Electronics(CENSE), Shanghai 200050, China}\affiliation{University
of Chinese Academy of Sciences, Beijing 100049, China}

\begin{abstract}
Low-temperature specific heat (SH) is measured on the 1111-type CaFe$_{0.88}$Co$_{0.12}$AsF single crystals
under different magnetic fields. A clear SH jump with the height $\Delta C/T|_{T_c}$ = 10.4 mJ/mol K$^2$ was observed at the superconducting transition temperature $T_c$.
The electronic SH coefficient $\Delta\gamma (B)$ increases linearly with the field below 5 T and a kink is observed around 5 T, indicating a multi-gap feature in the present system. 
Such a sign is also reflected in the $T_c-B$ data.
A detailed analysis shows that this behavior can be interpreted in terms of a two-gap scenario with the ratio $\Delta_{L}/\Delta_{S}= 2.8\sim4.5$. 

\end{abstract}

 \maketitle

Superconducting (SC) mechanism is the central issue in the study of unconventional superconductors. Since the discovery of Fe-based superconductors (FeSCs)~\cite{LaFeAsO}, many efforts have been made on this problem~\cite{Mazin2011}.
Gap structure can supply very important information for this issue, because typically different SC mechanism will predict distinct gap symmetry and structure. For example, the Fermi surfaces with a better nesting condition tend
to have a stronger pairing amplitude and larger SC gap in the itinerant mechanism~\cite{Mazin2009614,Graser2009}, while according to the local scenario, a larger SC gap should open on the smaller Fermi surface~\cite{JPHu2008}.
For the 1111 system with the ZrCuSiAs-type structure, the SC gap structure has been investigated by diverse methods and the conclusions are, however,
rather controversial. Most of the early studies based on the polycrystalline samples claimed a nodal gap structure~\cite{Mu2008,Matano2008,Shan2008}. Later on, the nodeless scenarios were also reported by other groups,
some of which were measured on single-crystalline samples~\cite{TYChen2008,Hashimoto2009,Malone2009}.
Overall, however, the investigations on this issue are still lacking. Especially, an in-depth specific heat study based on high-quality single crystals is almost blank, mainly because the SH measurements
usually require a considerable sample mass. Specific heat (SH) is one of the powerful tools to measure the quasiparticle density of states (DOS) at the Fermi level to
detect the information about the gap structure~\cite{Wen2004,Mu2007,Mu2009,Wen2009,Mu2010,Mu2011}.
The feature of the gap structure can be essentially determined by measuring the variation in the electronic SH versus temperature and magnetic field~\cite{review1,review2}. Obviously, more efforts are urgently required to obtain the
intrinsic thermodynamic property of this system.

Recently, due to the progresses on the single-crystal growth of the fluorine-based 1111 system Ca(Fe,Co)AsF~\cite{Ma2015,Ma2016}, systematic investigations
have been carried out on this system~\cite{Taichi2018,Xiao2016,Xiao2016-2,CaFeAsFCo-Hc2,Xu2018,Ma2018,Gao2018,Mu2018,Wang2019,Yu2019}. In our previous works, a two-gap feature is revealed by the temperature dependence of
lower critical field $H_{c1}(T)$ and point-contact spectroscopy measurements on the single-crystalline samples~\cite{Wang2019,Yu2019}. However, the ratio of the two gaps ($\Delta_{L}/\Delta_{S}$) 
is not very consistent between different measuring means. According to the data of $H_{c1}(T)$, $\Delta_{L}/\Delta_{S} \approx$ 5.2~\cite{Wang2019}. While from point-contact spectroscopy measurement, the ratio
$\Delta_{L}/\Delta_{S} \approx$ 2.6~\cite{Yu2019} is obtained. Thus more experiments are required to further clarify this issue.
In this paper, low temperature specific heat was measured on the 1111-type single crystals of CaFe$_{0.88}$Co$_{0.12}$AsF. A linear field dependence of the electronic SH coefficient $\Delta\gamma(B)$ is discovered in the
field region below 5 T, which turns to another linear evolution with a smaller slope under the even higher field. 
This is very like the behavior of MgB$_2$~\cite{Bouquet2002} and implies a clear two-gap picture. The ratio of the two gaps is estimated to be  $\Delta_{L}/\Delta_{S}= 2.8\sim4.5$.

The CaFe$_{0.88}$Co$_{0.12}$AsF single crystals were
grown by the self-flux method. The detailed growth conditions and sample characterizations has been reported in our previous work~\cite{Ma2015,Ma2016}.
In order to ensure a sufficient mass for the SH measurement, three high-quality single crystals with very similar magnetization transitions (see the inset of Fig. 1)
were chosen. The total mass of the three samples is 1.4 mg. The dc magnetization
measurements were done with a superconducting quantum interference
device (Quantum Design, MPMS 3). The specific heat were measured on the physical property measurement system
(Quantum Design, PPMS). We employed the thermal relaxation technique
to perform the specific heat measurements. The external
field was applied along the $c$ axis of all the three single
crystals during the SH measurements.

\begin{figure}
\includegraphics[width=8.5cm]{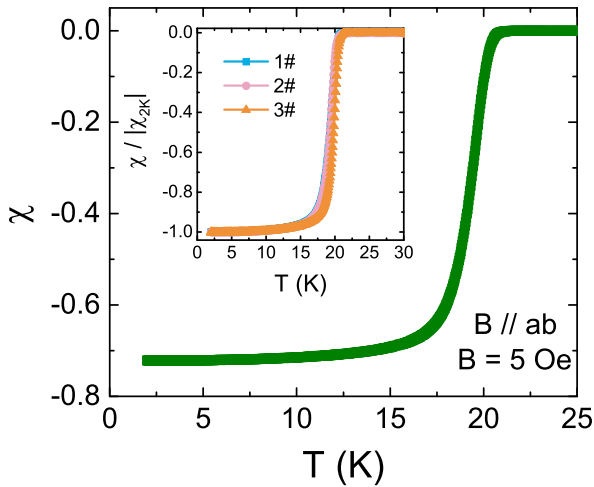}
\caption {(color online) Temperature dependence of total dc
magnetization for three CaFe$_{0.88}$Co$_{0.12}$AsF single crystals. The data is measured using the zero-field-cooling mode. The magnetic field of 5 Oe is applied along the $ab$ plane. 
The inset shows the data of the three samples respectively. The data is normalized to the absolute value at 2 K to have a clear comparison. } \label{fig1}
\end{figure}

\begin{figure}
\includegraphics[width=8cm]{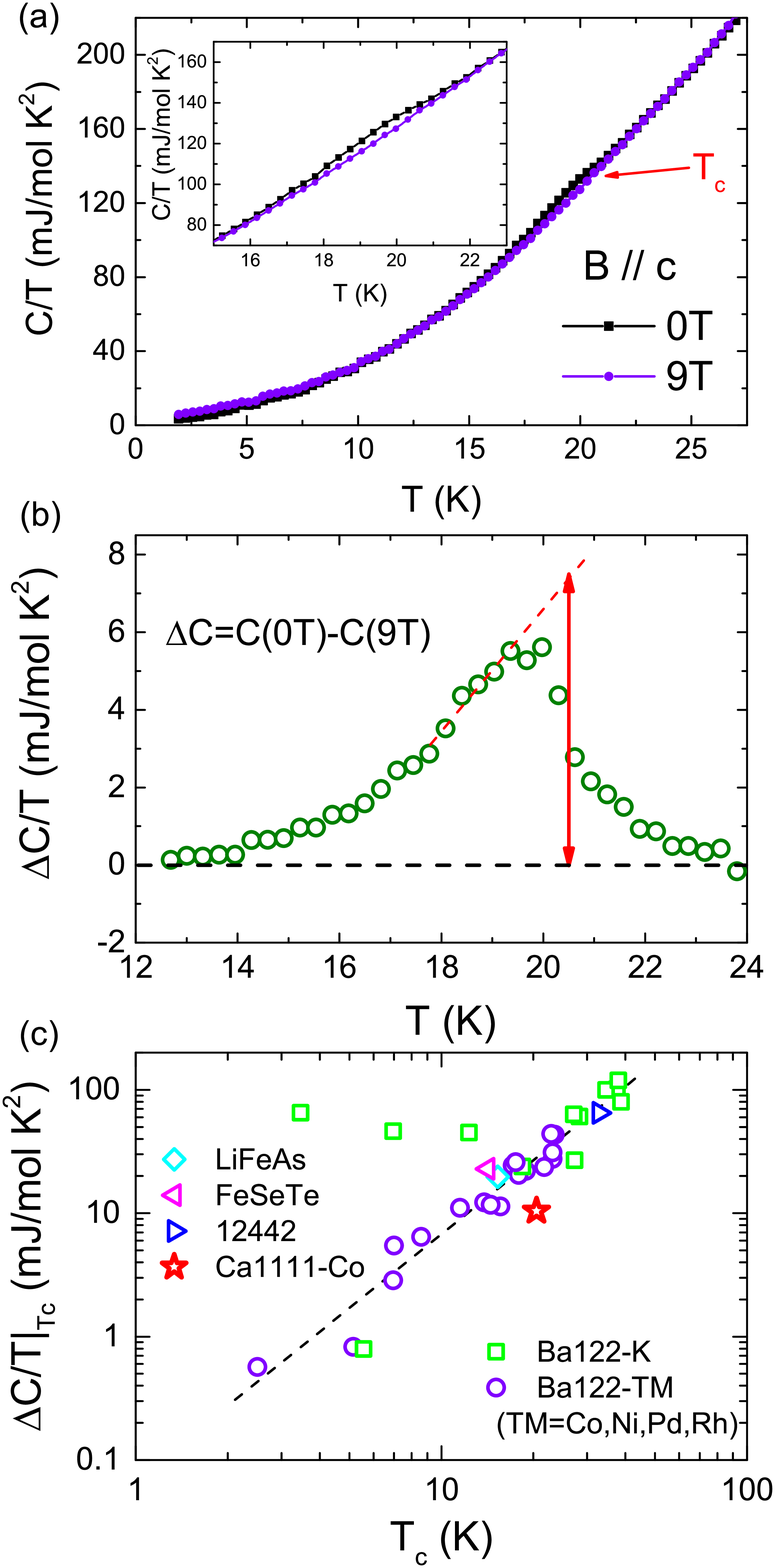}
\caption {(color online) (a) Temperature dependence of specific heat plotted as $C/T$ vs $T$ under two fields 0 T and 9 T. The inset shows an enlarged view of the data near the superconducting transition. 
(b) The SH anomaly near $T_c$ by subtracting the the data under 9 T from that under 0 T. 
(c) SH jump height $\Delta C/T|_{T_c}$ as a function of $T_c$. The result of CaFe$_{0.88}$Co$_{0.12}$AsF (Ca1111-Co, red star) is from the present work.
The data of other systems were collected from the references~\cite{BNC,BaCo2009,BaRh2009,Mu2010,BaK2008,BaK2009,Mu2009,BaK2013,Wang2019}.} \label{fig2}
\end{figure}

The superconducting transition of the
single crystals was checked by the dc magnetization measurements down to 1.8 K. In the inset of Fig. 1, we show the temperature dependence of the normalized magnetization data for three CaFe$_{0.88}$Co$_{0.12}$AsF samples.
The data were collected using a zero-field-cooling mode under a field of 5 Oe. The SC transitions for three samples are basically the same with the onset transition temperature $T_c$ = 21 K.
In the main frame of Fig. 1, we show the magnetic susceptibility of the three samples measured together. The magnetic field was applied parallel to the ab-plane of the crystal to minimize
the effect of the demagnetization when estimating the SC volume fraction. The SC volume fraction is determined to be about 72\%. Although this value is not very high,
the rather sharp SC transition indicates that the SC part of the sample is homogeneous and the discussion in this work is credible.

In Fig. 2(a) we show the raw data of SH coefficient $C/T$ vs $T$ under two differnet magnetic fields 0 T
and 9 T. In order to have a convenient comparison with the 122 system, here one mole means Avogadro number of unit cells or two times of formula units, [CaFe$_{0.88}$Co$_{0.12}$AsF]$_2$.
An unobvious SH anomaly can be seen from the raw SH data under zero field near $T_c$. The inset of Fig. 2(a) shows an enlarged view of this anomaly near the SC transition.
Under 9 T, the SH anomaly was suppressed markedly due to the field-induced pair-breaking effect and could not be distinguished from the raw data which include a large contribution from phonon.
In order to highlight the SH anomaly under 0 T, we subtracted the data under 9 T from the zero field data to eliminate influences of the phonon contributions and the results is shown in Fig. 2(b).
The SH anomaly at zero field was determined to be about $\Delta C/T|_{T_c,0}$ = 7.5 mJ/mol K$^2$, as indicated by the red arrowed line in Fig. 2(b). Considering the non-SC fraction of about 28\%, 
the actual magnitude of the SH anomaly should be $\Delta C/T|_{T_c,actual}$ = 10.4 mJ/mol K$^2$.
We found that this magnitude is clearly smaller than that observed in the 122 system with similar $T_c$~\cite{Mu2010} and markedly deviates from the plot based on
the Bud'ko, Ni and Canfield (BNC) law~\cite{BNC}, $\Delta C \propto T_c^3$ (see Fig. 2(c)). To have a meaningful comparison, the SH data for all the samples in this figure has been normalized to per [FeAs]$_2$.
Assuming a weak-coupling BCS picture where the ratio $\Delta C/\gamma_n T|_{T_c}$= 1.43, we can estimated the
normal state electronic SH coefficient $\gamma_n \approx$ 7.3 mJ/mol K$^2$. We note here that this value may be an upper limit of $\gamma_n$ since typically the coupling strength is larger than the BCS prediction of 1.43.

\begin{figure}
\includegraphics[width=9.5cm]{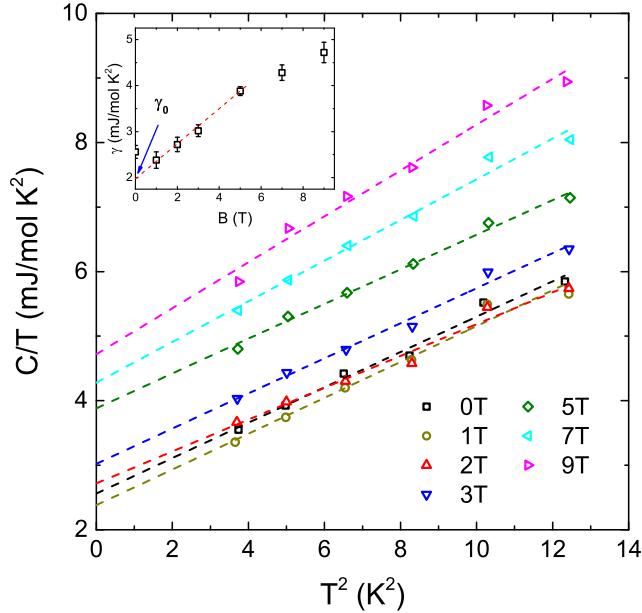}
\caption {(color online) The raw data of the SH under different
fields in the low temperature region. The data are shown in $C/T$ vs
$T^2$ plot. The dashed straight lines are the guides for the eyes. The inset shows the field dependence of the electronic SH coefficient $\gamma (B)$.}
\label{fig3}
\end{figure}

We next focus our attention on the SH data in the low temperature range
to study the low-energy quasiparticle (QP) excitations. The raw data of the SH in various magnetic fields in the low-temperature region below 3.5 K is shown as $C/T$ vs $T^2$ in Fig. 3.
The curves display a clear linear tendency in this low temperature region. No Schottky
anomaly can be observed, which facilitates the following analysis of our data. The SH data generally shows a monotonous upward shift with the increase of the magnetic field,
implying the QP excitation induced by the magnetic field.
By extrapolating this linear tendency to zero temperature (as shown by the dashed lines in Fig. 3), we can obtain the field dependence of the electronic SH coefficient $\gamma (B)$ because the
phonon contribution vanishes as the temperature is reduced to 0 K. As shown in the inset of Fig. 3, a residual term $\gamma_0 \equiv \gamma (0)\approx$ 2.0 mJ/mol K$^2$ was revealed. Considering the
rather small values of $\Delta C/T|_{T_c}$ and $\gamma_n$, the magnitude of $\gamma_0$ should not be ignored. Typically this term was
attributed to the non-superconducting fraction of the sample and/or the residual quasiparticle DOS in the SC materials with d-wave or S$^\pm$ gap symmetry~\cite{Wen2004,Mu2010,Bang2010}.
Since a SC volume fraction was determined to be above 72\% by the magnetization measurement (see Fig. 1), $\gamma_0$ ($\gamma_0/\gamma_n\approx$27\%) can mainly be attributed to the non-superconducting fraction of the sample.

The field-induced term $\Delta\gamma (B)=(\gamma (B)-\gamma_0$)/72\% is shown in Fig. 4(a). The data is divided by 72\% to deduce the intrinsic SC property of this material. 
$\Delta\gamma (B)$ increases linearly as the magnetic field increases from 0 T to 5 T 
and shows a kink feature around 5 T, above which another linear evolution can be seen with a smaller slope. 
This behavior is very similar to that observed in the famous multi-band superconductor MgB$_2$, where a two-gap picture is proposed~\cite{Bouquet2002}. 
Obviously such an observation is consistent qualitatively with the reported of the lower critical field and point-contact spectroscopy measurements~\cite{Wang2019,Yu2019}. 
In order to give a more precise understanding, we attempted to obtain the normal state values of $\gamma_n$ and
the out-of-plane upper critical field $B_{c2}$ ($\equiv B_{c2}(0)$). The value of $\gamma_n$ = 7.3 mJ/mol K$^2$ has been derived from the height of SH jump. As for the estimation of $B_{c2}$, 
experiments under high magnetic fields
reveal that the $B_{c2}-T$ curve shows a roughly linear behavior at low temperature~\cite{Fang2010,Khim2011} and doesn't display a flattening tendency near 0 K as predicted by the Werthamer-Helfand-Hohenberg (WHH) relation\cite{Werthamer1966}.
From the data of our previous work~\cite{CaFeAsFCo-Hc2}, we can obtain the $B_{c2}-T$ data of the present system near $T_c$ and replot it as $T_c$ vs $B$, as shown in Fig. 4(a) (right). 
Assuming a similar linear evolution of $B_{c2}(T)$ at low temperature, just as the high-field experiments have revealed, a zero-temperature value of $B_{c2}$ can be estimated to be 25 T (see the blue dashed line in Fig. 4(a)).
We note that this value does not exceed the the Pauli paramagnetic limit $B_P = 1.84 \times T_c$ = 39 T~\cite{ParaLimit}.
The position of the normal state ($B_{c2}$, $\gamma_n$) is shown by the red diamond in Fig. 4(a). 

\begin{figure}
\includegraphics[width=8.5cm]{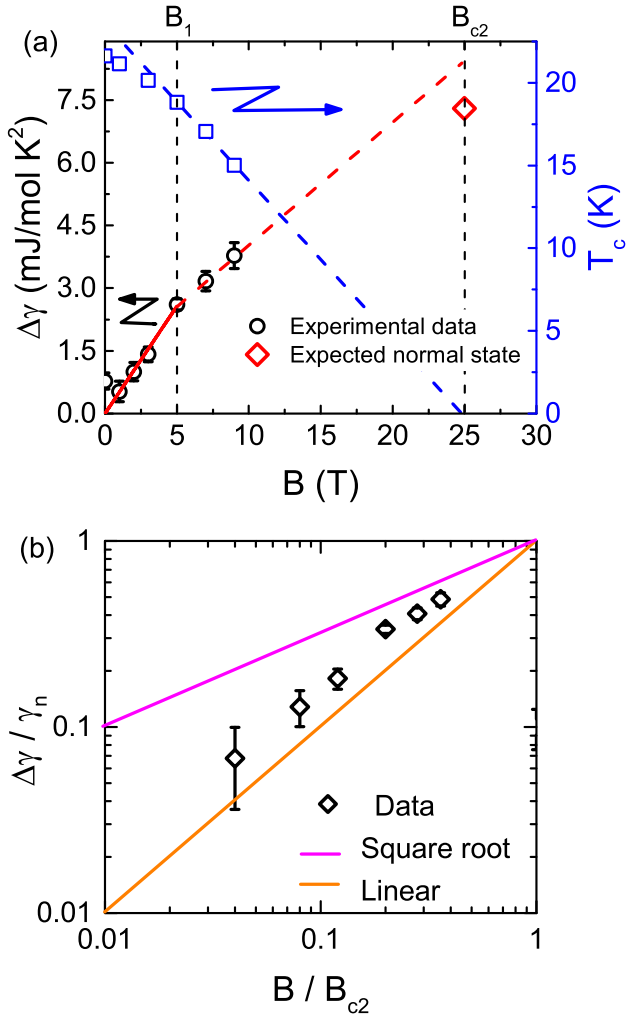}
\caption {(color online) (a) Field dependence of the of the electronic SH coefficient $\Delta\gamma$ (left) and $T_c$ (right). The solid and dashed lines are guides for eyes.
(b) The same data as (a) plotted in logarithmic scale. $\Delta\gamma$ and $B$ are normalized by $\gamma_n$ and $B_{c2}$ respectively. } \label{fig4}
\end{figure}

Following the tendency of the linear dependence above 5 T as shown by the red dashed line in Fig. 4(a), $\Delta\gamma (B)$ achieves the value slightly higher than the expected normal state value $\gamma_n$ at $B_{c2}$. 
This is reasonable because considerable overlapping of zero-energy density of states at each core can
make the linear behavior in $\Delta\gamma (B)$ change to nonlinear even for the isotropic gap~\cite{Nakai2004}. Moreover, a change of the slope can also be observed around $B_1$ = 5 T in the $T_c-B$ curve. 
Such an accordance supplies a good example that the multi-gap feature can be reflected
in the temperature dependent $B_{c2}(T)$ data.
As shown by the vertical dashed lines in Fig. 3(a), now we have two characteristic fields: $B_1$ and $B_{c2}$. Within the multiple-gap picture, 
these two fields corresponds to two gaps in different Fermi surfaces (FSs)~\cite{Bouquet2002,Bourgeois2016}. $B_1$ is a virtual upper critical field for band with a smaller gap while $B_{c2}$ 
is the upper critical field for that with a lager gap. In this case, we have $B_{c2}\sim(\Delta/v_F)^2$ and $\Delta_L/\Delta_S = (v_{F,L}/v_{F,S}) (B_{c2}/B_1)^{1/2}$, where $v_{F,L}$ and $v_{F,S}$ are the Fermi velocities
in the Fermi surface with the larger and smaller gap respectively. According our previous estimation~\cite{Wang2019}, we have $v_{F,L}/v_{F,S}$ = $1.25\sim2$. Considering the fact that $B_{c2}/B_1$ = 5, 
we can give an estimation $\Delta_L/\Delta_S$ = $2.8\sim4.5$.
To have a vivid impression, we replotted the $\Delta\gamma(B)$ data in logarithmic scale with both
coordinate axes normalized and showed the results in Fig. 4(b). All the experimental data locate in between the magenta and orange
lines, which represent the square root and linear relations between $\Delta\gamma/\gamma_n$ and $B/B_{c2}$ respectively. The square root behavior $\Delta\gamma/\gamma_n \propto \sqrt{B/B_{c2}}$ is a characteristic of
the SC gap with line nodes~\cite{Volovik-1,Volovik-2} and the linear relation is a consequence of the isotropic gap structure~\cite{Nakai2004}. 
Thus the degree of the gap anisotropy in the present system is lower than the line nodal case.

In summary, we studied the low-temperature specific heat of the 1111-type CaFe$_{0.88}$Co$_{0.12}$AsF single crystals. We found an SH jump with the height of 10.4 mJ/mol K$^2$, which diverges from the BNC scaling.
The electronic SH coefficient $\Delta\gamma$ shows a linear increase with field in the field region below 5 T and changes the slope with further increasing the field,
indicating a multi-gap behavior. The degree of the anisotropy was estimated to be $\Delta_{L}/\Delta_{S}=2.8\pm4.5$.

\begin{acknowledgments}
This work is supported by the Youth Innovation Promotion Association of the Chinese
Academy of Sciences (No. 2015187), the Natural Science Foundation of China
(No. 11204338), and the ``Strategic Priority Research Program (B)" of
the Chinese Academy of Sciences (No. XDB04040300). W.L. also acknowledges the start-up funding from Fudan University.
\end{acknowledgments}

\bibliography{CaFeCoAsF-SH}

\end{document}